# RELAXATION OF THE CHIRAL CHEMICAL POTENTIAL IN THE DENSE MATTER OF A NEUTRON STAR

**Maxim Dvornikov**[1,2*]

[1] *Pushkov Institute of Terrestrial Magnetism, Ionosphere, and Radio Wave Propagation, Moscow, Russia*
[2]*National Research Tomsk State University, Tomsk, Russia.*

*A model of the generation of a magnetic field in a neutron star is developed, based on an instability of the magnetic field caused by the electroweak interaction between electrons and nucleons in nuclear matter. The rate of change of the helicity of the electrons as they scatter on protons in the dense matter of a neutron star is calculated with the help of methods of quantum field theory. The influence of the electroweak interaction between electrons and background nucleons on the process of change of the helicity is examined. A kinetic equation is derived for the evolution of the chiral chemical potential. The results obtained are used to describe the evolution of the magnetic field in magnetars.*

## INTRODUCTION

Some neutron stars can possess extremely powerful magnetic fields $B \geq 10^{15}$ G. Such neutron stars are called magnetars [1]. Despite the long history of observations of magnetars and the existence of numerous theoretical models of the generation of their magnetic fields, at the present time there does not exist a generally accepted mechanism explaining the origin of the magnetic field in these compact stars. Recently, in [2, 3], efforts were undertaken to solve the problem of the generation of the magnetic field in magnetars with the help of the chiral magnetic effect. The given effect was used to explain the origin of toroidal magnetic fields in neutron stars in [4]. An alternative mechanism predicting the creation of strong cosmic magnetic fields based on the instability of the magnetic field caused by the interaction violating spatial parity was proposed in [5].

A new model of the generation of magnetic fields in magnetars was proposed in [6–8]. The main mechanism underlying the proposed model is the growth of a seed magnetic field due to its instability in nuclear matter caused by the electron-nucleon ($eN$) electroweak interaction. Within the framework of the proposed approach, some observed characteristics of magnetars have been explained, such as the strength and the length scale of the magnetic field, and also the age of these stars. Nevertheless, some aspects of the model require a more thorough grounding on the basis of calculations making use of methods of quantum field theory (QFT). The present paper is dedicated to a further development of the proposed description of the generation of magnetic fields in magnetars.

## 1. MODEL OF THE GENERATION OF MAGNETIC FIELDS IN MAGNETARS

As is well known, the dense matter of a neutron star consists of ultra-relativistic electrons and non-relativistic nucleons: neutrons and protons. It is assumed that the given matter possesses zero macroscopic velocity and zero polarization. Electrons in such matter interact with nucleons through the electroweak forces, which violate parity. In [6, 7] it was found that in this case, in an external magnetic field $\boldsymbol{B}$, an induced anomalous electric current of electrons $\boldsymbol{J}_5$ arises having the following form:

$$\boldsymbol{J}_5 = \Pi \boldsymbol{B}, \quad \Pi = \frac{2\alpha_{\text{em}}}{\pi}(\mu_5 + V_5), \tag{1}$$

* E-mail: maxdvo@izmiran.ru

where $\alpha_{\rm em} \approx 7.3 \cdot 10^{-3}$ is the fine structure constant, $\mu_5 = (\mu_{\rm R} - \mu_{\rm L})/2$ is the chiral chemical potential, $\mu_{\rm R,L}$ are the chemical potentials of the right and left electrons, $V_5 = (V_{\rm L} - V_{\rm R})/2 \approx G_{\rm F} n_n / 2\sqrt{2}$, $V_{\rm L,R}$ are the effective interaction potentials of the left and right electrons with the nucleons of the medium (mainly with neutrons), $G_{\rm F} \approx 1.17 \cdot 10^{-5}$ GeV$^{-2}$ is the Fermi constant, and $n_n$ is the neutron density. The current in equation (1) was derived in [6, 7] using the exact solution of the Dirac equation for an ultra-relativistic electron interacting with matter under the influence of an external magnetic field.

Basing ourselves on expression (1) for the current, in [7] we obtained a system of equations for the evolution of the spectrum of the density of the magnetic helicity $h(k,t)$, the spectrum of the density of the magnetic energy $\rho_{\rm B}(k,t)$, and the chiral chemical potential in the form

$$\frac{\partial h(k,t)}{\partial t} = -\frac{2k^2}{\sigma_{\rm cond}} h(k,t) + \frac{8\alpha_{\rm em}\left[\mu_5(t) + V_5\right]}{\pi \sigma_{\rm cond}} \rho_{\rm B}(k,t), \tag{2}$$

$$\frac{\partial \rho_{\rm B}(k,t)}{\partial t} = -\frac{2k^2}{\sigma_{\rm cond}} \rho_{\rm B}(k,t) + \frac{2\alpha_{\rm em}\left[\mu_5(t) + V_5\right]}{\pi \sigma_{\rm cond}} k^2 h(k,t), \tag{3}$$

$$\frac{{\rm d}\mu_5(t)}{{\rm d}t} = \frac{\pi \alpha_{\rm em}}{\mu_e^2 \sigma_{\rm cond}} \int {\rm d}k \left\{ k^2 h(k,t) - \frac{4\alpha_{\rm em}}{\pi}\left[\mu_5(t) + V_5\right] \rho_{\rm B}(k,t) \right\} - \Gamma_f \mu_5(t), \tag{4}$$

where $\Gamma_f$ is the rate of change of the helicity of the electrons in electron-proton ($ep$) collisions (see Section 2), and $\mu_e$ is the average chemical potential of the electron gas. The functions $h(k,t)$ and $\rho_{\rm B}(k,t)$ are related to the total magnetic helicity $H(t)$ and the magnetic field strength via the relations

$$H(t) = V \int h(k,t) {\rm d}k, \quad \frac{1}{2} B^2(t) = \int \rho_{\rm B}(k,t) {\rm d}k, \tag{5}$$

where $V$ is the normalization volume. The integration in formula (5) is performed over the entire range of the variation of the wave number $k$. It is necessary to point out that formula (5) assumes isotropic spectra.

Equations (2) and (3) for $h(k,t)$ and $\rho_{\rm B}(k,t)$ are a direct consequence of the modified Faraday equation, in which the anomalous current in relation (1) is taken into account. The first two terms on the right-hand side of Eq. (4) for $\mu_5(t)$ result from the Adler anomaly for ultra-relativistic electrons and from Eq. (2). The last term on the right-hand side of Eq. (4), $\Gamma_f \mu_5$, has been taken into account phenomenologically. It is based on the fact that the helicity of an electron changes in $ep$ collisions. As a rule, electrons are ultra-relativistic in neutron stars. However, they have nonzero mass. Thus, the following estimate was used in [6] for $\Gamma_f$:

$$\Gamma_f \sim \left(\frac{m}{\mu_e}\right)^2 \nu_{\rm coll} \sim \left(\frac{m}{\mu_e}\right)^2 \frac{\omega_p^2}{\sigma_{\rm cond}}, \tag{6}$$

where $m$ is the mass of an electron, $\nu_{\rm coll}$ is the frequency of the $ep$ collisions, and $\omega_p$ is the plasma frequency in a degenerate plasma. Equation (6) was derived on the basis of the relation between $\nu_{\rm coll}$ and $\sigma_{\rm cond}$ in a classical Lorentz plasma [9, pp. 66–67].

Thus, in order to complete the theoretical grounding of the main equations of the model in [6–8], it is necessary to consider the variation of the helicity of ~~the~~ electrons in $ep$ collisions in the dense matter of a neutron star using methods of QFT. Moreover, it is not without interest to investigate the influence of the electroweak interaction between ~~the~~ electrons and ~~the~~ nucleons on the given process.

## 2. ELECTRON-PROTON COLLISIONS IN A DENSE PLASMA

In a neutron star, the helicity of a massive electron can vary during $ep$ and electron-electron collisions due to the electromagnetic interaction with exchange of a virtual plasmon, and also upon the interaction of an electron with the anomalous magnetic moment of the neutron. In [10] it was found that the frequency of $ep$ collisions in the dense matter of a neutron star is much higher than for other reactions. Thus, only $ep$ collisions need be taken into account.

### 2.1. Rate of change of helicity in $ep$ collisions

The matrix element of $ep$-scattering due to the electromagnetic interaction has the form

$$\mathcal{M} = \frac{\mathrm{i}e^2}{\left(k_1 - k_2\right)^2} \bar{e}(p_2)\gamma^\mu e(p_1)\, \bar{p}(k_2)\gamma_\mu p(k_1), \tag{7}$$

where $e > 0$ is the absolute value of the charge of the electron, $\gamma^\mu = \left(\gamma^0, \gamma\right)$ are the Dirac matrices, $p^\mu_{1,2} = \left(E_{1,2}, \boldsymbol{p}_{1,2}\right)$ and $k^\mu_{1,2} = \left(E_{1,2}, \boldsymbol{k}_{1,2}\right)$ are the 4-momenta of the electrons and protons. The momenta of the particles before and after the scattering have the indices 1 and 2. Besides the exchange of a plasmon with the proton, the electron can interact via the electroweak interaction with the nucleons of the matter of a neutron star. In order to take this interaction into account, in matrix element (7), instead of solutions of the Dirac equation in a vacuum it is necessary to use the spinors corresponding to exact solutions of the wave equation for an electron interacting with the background matter. These exact solutions are found in [11].

We will be interested in reactions in which the helicity of the electron changes sign (flips). Let us first consider $e_\mathrm{R} \to e_\mathrm{L}$ transitions. According to Eq. (7), it is necessary to calculate the following quantity:

$$J^\mu = \left(J_0, \boldsymbol{J}\right) = \bar{u}_-(p_2)\gamma^\mu u_+(p_1), \tag{8}$$

where $u_\pm$ are the basis spinors corresponding to the different polarizations of the electrons. They are normalized by the condition $u^\dagger_\pm u_\pm = 1$. Direct calculation of $J^\mu$ gives

$$J_0 = -\frac{mP_0\left[p_1 + p_2 + E_+(p_1) + E_-(p_2) - 2\bar{V}\right]}{2\sqrt{E_{0+}(p_1)E_{0-}(p_2)\left[E_-(p_2) + p_2 - V_\mathrm{R}\right]\left[E_+(p_1) + p_1 - V_\mathrm{L}\right]}},$$

$$\boldsymbol{J} = -\frac{m\boldsymbol{P}\left[p_1 - p_2 + E_+(p_1) - E_-(p_2) - 2V_5\right]}{2\sqrt{E_{0+}(p_1)E_{0-}(p_2)\left[E_-(p_2) + p_2 - V_\mathrm{R}\right]\left[E_+(p_1) + p_1 - V_\mathrm{L}\right]}}, \tag{9}$$

where

$$P_0 = w_-^\dagger(\boldsymbol{p}_2) w_+(\boldsymbol{p}_1), \; \boldsymbol{P} = w_-^\dagger(\boldsymbol{p}_2)\boldsymbol{\sigma} w_+(\boldsymbol{p}_1). \tag{10}$$

Here $\boldsymbol{\sigma}$ are the Pauli matrices, $E_{0\pm}(p) = |p \mp V_5|$, $\bar{V} = \left(V_\mathrm{L} + V_\mathrm{R}\right)/2$, and $w_\pm(\boldsymbol{p})$ are the two-component spinors appearing on p. 86 in [12]. To derive formula (9), we have made use of the Dirac matrices in the chiral representation.

As was shown in [13, pp. 205–209], to describe the collisions in relativistic plasma due to the long-range Coulomb forces, it is necessary to use the elastic scattering approximation. Thus, in the treatment of the $\mathrm{R} \to \mathrm{L}$ transitions, it is necessary to assume that $E_+(p_1) = E_-(p_2)$. Using the expression for the energies of the ultra-relativistic electrons [11] $E_\pm(p_{1,2}) = p_{1,2} + V_{\mathrm{R,L}}$, we find that the condition for elasticity of the collisions is written in the form $p_1 - p_2 = 2V_5$. In this case, we find from relation (9) that $\boldsymbol{J} = 0$. The quantities $P_0$ in formula (10) can also be calculated in explicit form: $|P_0|^2 = \left[1 - (\boldsymbol{n}_1 \cdot \boldsymbol{n}_2)\right]/2$, where $\boldsymbol{n}_{1,2}$ are the unit vectors in the direction $\boldsymbol{p}_{1,2}$. Thus, employing formula (9), we find that the square of the matrix element given by formula (7) has the form

$$|\mathcal{M}|^2 = e^4 m^2 \frac{(p_1 + p_2)^2\left[1 - (\boldsymbol{n}_1 \cdot \boldsymbol{n}_2)\right]}{8(p_1 - V_5)^2 (p_2 + V_5)^2} \frac{\left[\mathcal{E}_1\mathcal{E}_2 + M^2 + (\boldsymbol{k}_1 \cdot \boldsymbol{k}_2)\right]}{\left[(\mathcal{E}_1 - \mathcal{E}_2)^2 - (\boldsymbol{k}_1 - \boldsymbol{k}_2)^2\right]^2}, \tag{11}$$

where the leading order in the mass of the electron is retained. Note that the contribution of the protons to $|\mathcal{M}|^2$, which are assumed to be non-polarized, is found according to the standard scheme (see, for example, [12, pp. 252–256]).

The total scattering probability has the form [12, pp. 248–249]

$$W = \frac{V}{2(2\pi)^8} \int \frac{\mathrm{d}^3 p_1 \mathrm{d}^3 p_2 \mathrm{d}^3 k_1 \mathrm{d}^3 k_2}{\mathcal{E}_1 \mathcal{E}_2} \delta^4(p_1 + k_1 - p_2 - k_2) |\mathcal{M}|^2$$

$$\times f_e(E_1^+ - \mu_\mathrm{R})\left[1 - f_e(E_2^- - \mu_\mathrm{L})\right] f_p(\mathrm{E}_1 - \mu_p)\left[1 - f_p(\mathrm{E}_2 - \mu_p)\right], \tag{12}$$

where we have carried out the summation over the polarizations of the proton after scattering. Here $f_{e,p}(E) = \left[\exp(\beta E) + 1\right]^{-1}$ are the Fermi–Dirac distributions of the electrons and protons, $\beta = 1/T$ is the inverse temperature, and $\mu_p$ is the chemical potential of the protons. In Eq. (12) it is assumed that electrons before and after a collision have different chemical potentials: $\mu_\mathrm{R}$ and $\mu_\mathrm{L}$. The protons and electrons are assumed to be in thermal equilibrium with the same temperature $T$.

Direct calculation of the integrals over the momenta of the ultra-relativistic electrons and non-relativistic protons in formula (12) gives

$$W(R \to L) = W_0(\mu_\mathrm{R} - \mu_\mathrm{L})\theta(\mu_\mathrm{R} - \mu_\mathrm{L}), \quad W_0 = \frac{Ve^4}{32\pi^5} \frac{m^2 M}{\mu_e} T \left[\ln\left(\frac{48\pi}{\alpha_\mathrm{em}}\right) - 4\right], \tag{13}$$

where $M$ is the mass of a proton. Note that in the derivation of expression (13), we took the dependence on the interaction potentials of the electrons with matter $V_{\mathrm{L,R}}$ exactly into account. In the treatment of the $L \to R$ transitions, the calculation of the total scattering probability is analogous to the $\mathrm{R} \to \mathrm{L}$ case. It can be shown that $W(\mathrm{L} \to \mathrm{R})$ in this situation coincides completely with formula (13) with the substitution $\mu_R \leftrightarrow \mu_L$ taken into account. For the sake of brevity, the corresponding calculations are not presented.

### 2.2. Kinetics of the chiral chemical potential

Proceeding from formula (13) and the analogous relation for the $L \to R$ transitions, we obtain the kinetic equations for the total number of right and left electrons $N_{R,L}$ in the form

$$\frac{dN_R}{dt} = -W(R \to L) + W(L \to R) = -W_0(\mu_R - \mu_L), \quad \frac{dN_L}{dt} = -W(L \to R) + W(R \to L) = -W_0(\mu_L - \mu_R). \quad (14)$$

Introducing the number densities of left and right electrons $n_{R,L} = N_{R,L}/V$ and employing the expression for $n_{R,L}$ in terms of the distribution function

$$n_{R,L} = 2\int \frac{d^3p}{(2\pi)^3} \frac{1}{\exp\left[\beta\left(p + V_{R,L} - \mu_{R,L}\right)\right] + 1} \approx \frac{\left(\mu_{R,L} - V_{R,L}\right)^3}{3\pi^2}, \quad (15)$$

we obtain that $d\left(n_R - n_L\right)/dt \approx 2\dot{\mu}_5 \mu_e^2/\pi^2$, where it has been taken into account that $\dot{V}_5 = 0$ and $\mu_5 << \mu_e$. Finally, we can derive the kinetic equation for $\mu_5$:

$$\frac{d\mu_5}{dt} = -\Gamma_f \mu_5, \quad \Gamma_f = \frac{\alpha_{em}^2}{\pi}\left[\ln\left(\frac{48\pi}{\alpha_{em}}\right) - 4\right]\left(\frac{m}{\mu_e}\right)^2 \left(\frac{M}{\mu_e}\right) T. \quad (16)$$

Here we have made use of formulas (13) and (14).

It should be noted that the quantity $\Gamma_f$ in formula (6) differs from that used in [6–8]. The reason for the difference in $\Gamma_f$ in formulas (6) and (16) consists in the fact that in [6] we used the results of [10], which considered scattering of non-polarized electrons on protons. However, in the problem considered in the present work, fixed polarizations of electrons, which have opposite values before and after scattering, are important. This fact explains, for example, the result that $\Gamma_f$ in formula (16) is linear in $T$, whereas in relation (6) the given quantity $\sim T^2$.

Note also that $\Gamma_f$ was recently calculated in [14]. The value of $\Gamma_f$ obtained in [14] does not depend on $T$, since in [14] it was assumed that the protons are non-degenerate. This assumption is valid in the early stages of the evolution of neutron stars. The present work examines the generation of a magnetic field in a neutron star found in a state of thermodynamic equilibrium at $t \geq 10^2$ years after the onset of the collapse of a supernova. At the given stage of the evolution of a neutron star, the proton component of matter is degenerate. It should be noted that $\Gamma_f \sim \alpha_{em}^2$ in formula (16), which coincides with the result of [14].

### 2.3. Thermodynamic description of relaxation of the chiral chemical potential

Recently, Sigl and Leite [2] advanced the hypothesis that the kinetics of the chiral chemical potential in the systems of left and right electrons interacting via the electroweak interaction with matter satisfies the equation

$$\frac{d\mu_5}{dt} = -\Gamma_f(\mu_5 + V_5), \tag{17}$$

and not Eq. (16), which follows from the results of our calculations. It is possible, nevertheless, to show that Eq. (17) contradicts the laws of thermodynamics. Bringing formula (15) to bear, it is possible to rewrite Eq. (17) in the form

$$\frac{d}{dt}(n_R - n_L) = -\frac{2\mu_e^2}{\pi^2}\Gamma_f(\mu_5 + V_5). \tag{18}$$

It is clear from formula (18) that the equilibrium state in which $n_{R,L} = \text{const}$ would be reached at $\tilde{\mu}_R = \tilde{\mu}_L$, where $\tilde{\mu}_{L,R} = \mu_{L,R} - V_{L,R}$, and not at $\mu_R = \mu_L$, as is required by the laws of thermodynamics [15, p. 306]. Note that the formally introduced quantities $\tilde{\mu}_{L,R} = \tilde{\mu}_{L,R}(P,T)$, where $P$ is the pressure in the system, are the chemical potentials in the absence of the background matter.

Analysis of the equilibrium state in a system of left and right electrons is a particular case of the description of the equilibrium of a body in an external field $\mathcal{V}(\boldsymbol{r})$. As was shown in [15, p. 73–74], the equilibrium in the given case is reached when the full chemical potential $\mu = \tilde{\mu}(P,T) + \mathcal{V}$ takes constant values inside the system. The results of [15, p. 73–74] are easily generalized to the case of a system consisting of two types of particles: left and right electrons. In this situation we find that in the equilibrium state the full chemical potentials, which include the interaction potentials with the background matter $V_{L,R}$, should coincide: $\mu_L = \mu_R$. Thus, Eq. (17), proposed in [2], not only is not confirmed by direct calculation of the probability of the processes $e_{L,R} \leftrightarrow e_{R,L}$ in Subsection 2.2, but also contradicts the conclusions of macroscopic thermodynamics.

### 3. GENERATION OF MAGNETIC FIELDS IN MAGNETARS

As the initial condition for the magnetic energy density spectrum, we choose the Kolmogorov spectrum: $\rho_B(k,t_0) = \mathcal{C}k^{-5/3}$, where the constant $\mathcal{C}$ is related to the seed field in a young pulsar $B(t_0) = B_0 = 10^{12}$ G by the second of Eqs. (5). The integration in Eq. (5) is taken over the interval $k_{min} < k < k_{max}$, where $k_{min} = 2 \cdot 10^{-11}\,\text{eV} = R_{NS}^{-1}$, $R_{NS} = 10$ km is the radius of the neutron star, $k_{max} = \Lambda_B^{-1}$, and $\Lambda_B$ is the minimal scale of the magnetic field and is a free parameter. The initial magnetic energy density spectrum is chosen in the form $h(k,t_0) = 2q\rho_B(k,t_0)/k$, where $0 \le q \le 1$ is a parameter defining the initial helicity: $q = 0$ corresponds to zero helicity, and $q = 1$ corresponds to the maximal helicity. We choose the initial value of the chiral chemical potential to be $\mu_5(t_0) = 1$ MeV. Note that the evolution of the magnetic field has practically no dependence on $\mu_5(t_0)$ because of the largeness of $\Gamma_f$.

The number densities of electrons, protons, and neutrons are assumed to be equal $n_e = n_p = 9 \cdot 10^{36}\,\text{cm}^{-3}$ and $n_n = 1.8 \cdot 10^{38}\,\text{cm}^{-3}$. This corresponds to $\mu_e = 125$ MeV for relativistic electrons. Matter with such parameters can be present in neutron stars. In order to account for the energy balance in the system consisting of the magnetic field and the background matter, it is necessary to renormalize the parameter $\Pi$ in formula (1) (see the Appendix):

$$\Pi \to \Pi\left(1 - B^2/B_{eq}^2\right), \tag{19}$$

where $B$ and $B_{\text{eq}}$ are assigned in formulas (5) and (21). Substitution (19) also makes it possible to eliminate the excessive growth of the magnetic field at $t >> t_0$. A renormalization analogous to that implied by substitution (19) for $B << B_{\text{eq}}$ was also implemented in [8].

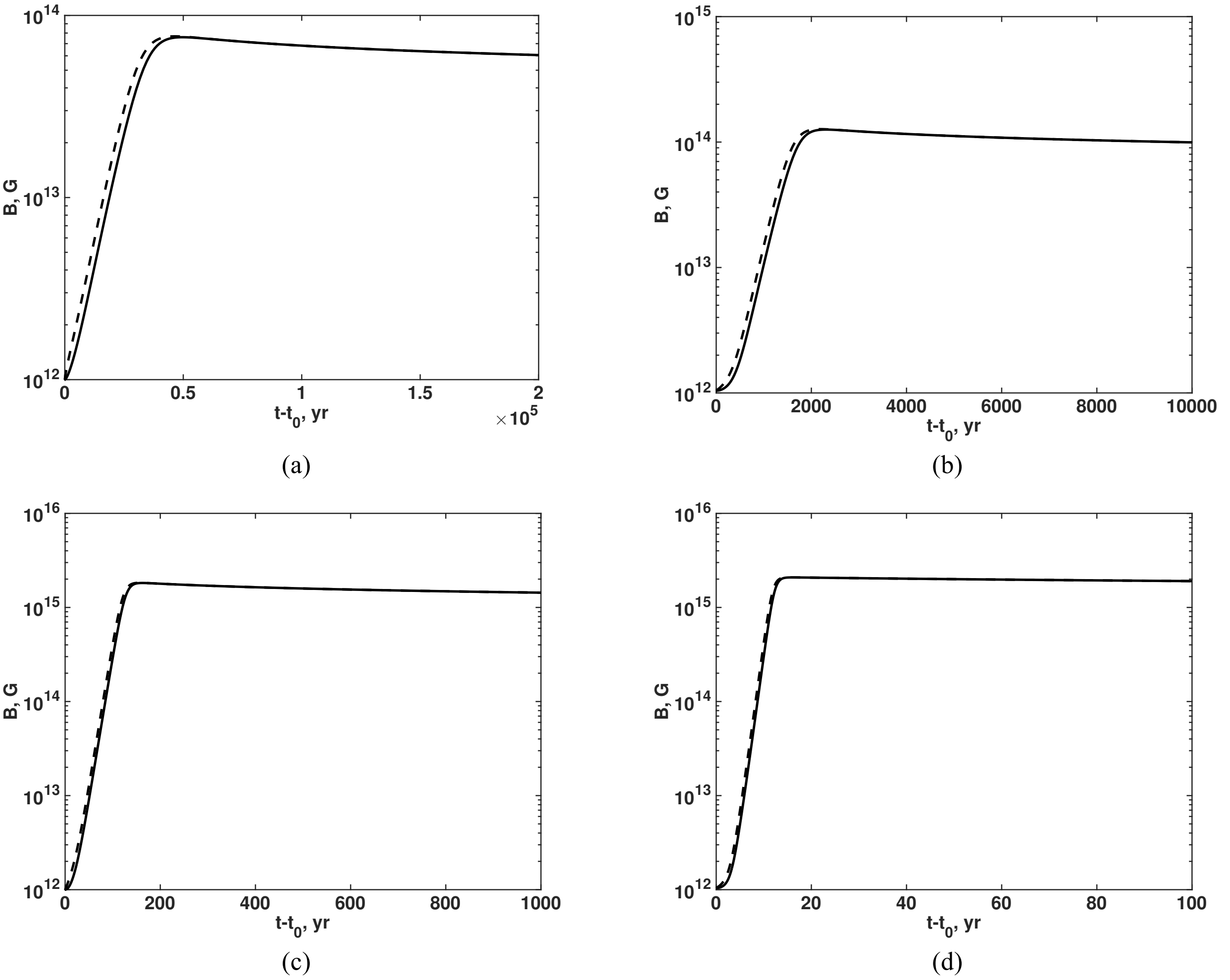

Fig. 1. Dependence of the magnetic field in a magnetar on $t - t_0$ for different values of $k_{\text{max}}$ and $T_0$. The solid curves correspond to magnetic fields with zero initial helicity ($q = 0$), and the dashed curves, to those with maximum initial helicity ($q = 1$): *a*) evolution of the magnetic field for $k_{\text{max}} = 2 \cdot 10^{-10}$ eV ($\Lambda_B = 1$ km) and $T_0 = 10^8$ K, *b*) growth of the magnetic field for $k_{\text{max}} = 2 \cdot 10^{-9}$ eV ($\Lambda_B = 10^2$ m) and $T_0 = 10^8$ K, *c*) evolution of the magnetic field for $k_{\text{max}} = 2 \cdot 10^{-10}$ eV ($\Lambda_B = 1$ km) and $T_0 = 10^9$ K, and *d*) growth of the magnetic field for $k_{\text{max}} = 2 \cdot 10^{-9}$ eV ($\Lambda_B = 10^2$ m) and $T_0 = 10^9$ K.

Let us consider the evolution of the magnetic field in a neutron star found in a state of thermal equilibrium that sets up for $t_0 < t \leq 10^6$ years, where $t_0 \sim 10^2$ years. In the given time interval, the neutron star cools down due to the emission of neutrinos in modified Urca processes. This leads to the following time dependence of the temperature [16]: $T(t) = T_0 \left(t/t_0\right)^{-1/6}$, where $T_0 = 10^8 - 10^9$ K is the temperature corresponding to $t = t_0$. Using the results of [10], it is possible to obtain the time dependence of the conductivity $\sigma_{\text{cond}}(t) = \sigma_0 \left(t/t_0\right)^{1/3}$, where $\sigma_0 = 2.7 \cdot 10^8 (T_0/10^8 \text{ K})^{-2}$ MeV is the conductivity at $t = t_0$. It is also necessary to take into account the new time dependence of $\Gamma_f$ in formula (16): $\Gamma_f = 1.6 \cdot 10^{11} \text{s}^{-1} (t/t_0)^{-1/6}$, where we have employed the cooling law of the neutron star and the chosen electron number density.

Fig. 1 plots the time dependence of the magnetic field strength based on a numerical solution of Eqs. (2)–(4) with the chosen initial conditions. It is clear from Fig. 1 that the magnetic field displays an exponential growth in the initial stage of the evolution. Such a growth of the field is mediated by the electroweak $eN$-interaction and is governed by the nonzero parameter $V_5$. The growth of the field takes place for $t - t_0 \sim (10 - 10^5)$ years, depending on $\Lambda_B$ and $T_0$. The most rapid growth is observed for $T_0 = 10^9$ K, which corresponds to the smallest value of $\sigma_{cond}$, and also for small-scale fields. Note that the time it takes the field to grow to its maximum value $t \sim (10^3 - 10^5)$ years for $T_0 = 10^8$ K (see Figs. 1*a* and *b*) is close to the observed age of young magnetars [1].

The maximum magnetic field strength $B_{max} \sim 10^{14} - 10^{15}$ G is determined by the initial thermal energy. After it is reached, the field begins slowly to decrease. This is explained by the continuing energy losses due to the emission of neutrinos. Note that $B_{max}$ in Fig. 1*a* is less than in Fig. 1*b*. This follows from the fact that the time scale of the field in Fig. 1*a* is greater and correspondingly ~~the~~ neutrinos have enough time to carry off more energy from the neutron star. It should be noted that the values of $B_{max} \sim 10^{15}$ G in Figs. 1*c* and *d* correspond to the predictions of magnetic fields in magnetars [1].

## 5. DISCUSSION OF RESULTS

The rate of change of the helicity of the electrons during $ep$ collisions was estimated in [6, 7] on the basis of qualitative arguments of classical physics [9, pp. 66–67]. As is well known, the spin of a particle is an intrinsically quantum object. It is namely for this reason that its evolution must be studied in a corresponding fashion. In the present work, we have implemented the methods of QFT to calculate the rate of the change of the helicity. This can explain the difference between the results obtained here and those of [6, 7] (see formulas (16) and (6)).

Another important result obtained in the given work consists in the findings of our study of the influence of the electroweak interaction of electrons with neutrons on the process of the change of their helicity during collisions with protons. Employing the method of exact solutions of the Dirac equation in an external field and assuming that the scattering is elastic and the electrons are ultra-relativistic, we found that the effective potentials $V_{L,R}$ do not enter into the expression for the total probability of the processes $e_{L,R} \leftrightarrow e_{R,L}$ in formula (13) in the explicit form. Hence it follows that the kinetic equation for the chiral chemical potential (Eq. (16)) coincides with the kinetic equation used in [6–8], in contradiction to the recent assertion otherwise in [2] (see Eq. (17)). Moreover, the kinetic equation derived here (Eq. (16)) is confirmed by the laws of thermodynamics (see Subsection 2.3).

Finally, an accurate account of the energy conservation law in the Appendix allowed us to alter the form of the quenching of the parameter $\Pi$ in formula (19) in comparison with the results of [8]. This led to a more correct description of the evolution of the magnetic field in magnetars in Section 3, especially for $B \sim B_{eq}$. Nevertheless, the parameters of the generated field, $B_{max} \sim 10^{14} - 10^{15}$ G, and the time required for the field to grow to its maximum value $\leq 5 \cdot 10^4$ years, are in agreement with astrophysical predictions for magnetars [1].

## APPENDIX. ENERGY SOURCE UNDERLYING GROWTH OF THE MAGNETIC FIELD

Despite the fact that the fermions in a neutron star are strongly degenerate, they possess a nonzero temperature. For example, $t \sim 10^2$ years after the explosion of a supernova the temperature may reach $T \gtrsim 10^8\,\mathrm{K}$. In [8] the hypothesis was advanced that the growth of the magnetic field predicted in [6, 7], can be supported by the transformation of the thermal energy of the fermions of matter into the energy of the magnetic field. In order to ground the possibility of the given process, it is necessary to consider the equation describing the conservation of energy in magnetohydrodynamics [17, pp. 226 – 227]:

$$\frac{\partial}{\partial t}\left(\frac{\rho \boldsymbol{v}^2}{2} + \varepsilon_T + \frac{B^2}{2}\right) = -(\nabla \cdot \boldsymbol{q}), \tag{20}$$

where $\rho$ is the mass per unit volume of the matter of a neutron star, $\varepsilon_T$ is the internal energy per unit volume, $\boldsymbol{v}$ is the velocity, and $\boldsymbol{q}$ is the energy flux density.

It is possible to represent $\varepsilon_T$ in the form $\varepsilon_T = \varepsilon_0 + \delta\varepsilon_T$, where $\varepsilon_0$ is the temperature-independent internal energy of the degenerate gas, and $\delta\varepsilon_T$ is the thermal correction. It was shown in [8] that the magnetic field can acquire energy from $\delta\varepsilon_T$. Moreover, it was found in [8] that

$$\delta\varepsilon_T = \frac{B_{\mathrm{eq}}^2}{2} = \left[\frac{M_{\mathrm{N}}\left(p_{\mathrm{F}_n} + p_{\mathrm{F}_p}\right)}{2} + \mu_e^2\right]\frac{T^2}{2}, \tag{21}$$

where $M_{\mathrm{N}}$ is the mass of the nucleons and $p_{\mathrm{F}_{p,n}}$ are the Fermi momenta of the protons and neutrons. Integrating Eq. (20) over the volume of the neutron star $V$ and assuming that $\boldsymbol{q} = 0$ on the surface of the neutron star, and also taking formula (21) into account, we obtain the conservation law

$$\frac{\mathrm{d}}{\mathrm{d}t}(\delta\varepsilon_T + \rho_{\mathrm{B}}) = 0, \quad \rho_{\mathrm{B}} = \frac{1}{2V}\int B^2 \mathrm{d}^3x. \tag{22}$$

Relation (22) shows that the growth of the magnetic field takes place owing to a decrease in the thermal correction to the internal energy. In Eq. (22) it was taken into account that $\dot{\varepsilon}_0 = 0$.

Bringing Eq. (22) to bear, it is possible to ground the quenching of the parameter $\Pi$ in formula (19). If we neglect cooling of the neutron star due to the emission of neutrinos, then, integrating Eq. (22) with the corresponding initial condition, it is possible to show that $B^2 + B_{\mathrm{eq}}^2(T) = B_{\mathrm{eq}}^2(T_0)$, where it has been taken into account that $B_{\mathrm{eq}}(T_0) >> B_0$ in a typical neutron star. Consequently, the temperature of matter in a neutron star will depend on the growing magnetic field as $T^2 = T_0^2[1 - B^2/B_{\mathrm{eq}}^2(T_0)]$, where we have utilized formula (21). Next, basing ourselves on the temperature dependence of the conductivity $\sigma_{\mathrm{cond}} >> 1/T^2$ [10], we find that the conductivity now depends on the growing magnetic field:

$$\sigma_{\mathrm{cond}} \to \sigma_{\mathrm{cond}}\left[1 - B^2 / B_{\mathrm{eq}}^2(T_0)\right]^{-1}. \tag{23}$$

Note that it is sufficient to take the given dependence into account in Eqs. (2)–(4) only in the terms containing $\mu_5 + V_5$ since only they are responsible for the instability of the magnetic field. If we also allow for cooling of the neutron star due to the emission of neutrinos, then it is necessary to make the substitution $B_{\text{eq}}^2(T_0) \to B_{\text{eq}}^2(T)$ in formula (23). The given modification of Eqs. (2)–(4) is equivalent to formula (19).

It is interesting to note that, despite the cooling of a neutron star due to the growth of its magnetic field, the second law of thermodynamics is not violated. This fact can be verified using the equation of the heat transfer in magnetohydrodynamics [17, p. 335], from which it follows that the total entropy of the neutron star always grows: $\dot{S} > 0$.

The author expresses his gratitude to V. G. Bagrov and V. B. Semikoz for fruitful discussions.

This work was supported by the Russian Foundation for Basic Research (Grant No. 15-02-00293) and the Program for Enhancement of Competitiveness of Tomsk State University among the World's Leading Research and Education Centers (Scientific-Research Work No. 8.1.01.2015).